\documentclass{phbauth}
\usepackage{graphicx}

\begin{document}

\begin{frontmatter}

\title{Spin-Gap States of a Periodic Mixed Spin Chain}

\author{Ken'ichi Takano\thanksref{thank1}}

\address{
Laboratory of Theoretical Condensed Matter Physics and 
Research Center for Advanced Photon Technology, \\
Toyota Technological Institute, Nagoya 468-8511, Japan}

\thanks[thank1]{E-mail: takano@toyota-ti.ac.jp}

\begin{abstract}
     We examine a chain of periodic arrays of 4 quantum spins 
with magnitudes of 1/2, 1, 3/2 and 1. 
     There are four kinds of nearest-neighbour exchange parameters 
among them. 
     We choose two independent parameters for concreteness: 
one represents the ratio of typical exchange parameters, and 
the other represents a distortion. 
     We determine the phase diagram of the ground state 
in the parameter space. 
     The phase boundaries appear as gapless lines 
which separate gapful disordered phases. 
     They are determined by the gapless equation which was 
previously derived by mapping a general periodic spin chain to 
the nonlinear $\sigma$ model. 
\end{abstract}

\end{frontmatter}

\section{Introduction}

      Since the discovery of cuprate superconductors, quantum 
spin systems have been studied to understand the magnetic 
properties of their undoped mother materials. 
      In particular quantum spin systems with spin gap are 
interesting since the spin gap is possibly related to the 
superconducting gap in the doped case. 
      Although the materials are two-dimensional, it is 
basically important to study magnetic properties of spin gap 
systems in any dimensions. 

      In one dimension, Haldane \cite{Haldane} predicted that 
the spin excitation spectrum of 
a uniform spin chain is gapless if the magnitude of a spin is 
a half-odd-integer, and is gapful if it is an integer. 
      The prediction has been confirmed theoretically and 
experimentally in many standpoints. 
      Since Haldane's prediction is based on a mapping of a 
spin Hamiltonian to the nonlinear $\sigma$ model (NLSM), 
it is recognized that the NLSM is a useful tool to investigate 
spin systems. 

      The NLSM has been developed for periodic inhomogeneous 
spin chains. 
      Affleck \cite{Affleck1} derived and examined 
an NLSM for a spin chain with bond alternation. 
      The application of the NLSM to spin chains with more than 
one spin species in the period of more than two lattice spacings 
are also considered \cite{Fukui,Takano1}. 
      In particular we have succeeded to derive an NLSM, with 
accurately keeping the degrees of freedom of the spin variables, 
for a general mixed spin chain with arbitrary finite period 
\cite{Takano1}. 

      We applied the NLSM method to systems with period 4 and 
spin species 2 (we defines the magnitudes as $s_a$ and $s_b$)
\cite{Takano2,Takano3}. 
      Imposing the condition that the ground state is singlet, 
only the possible order of spin magnitudes in a unit cell is 
of $s_a$-$s_a$-$s_b$-$s_b$. 
      We obtained phase diagrams in the parameter space of 
the exchange couplings. 
      In a few special cases, numerical calculations have been 
performed \cite{Chen,Tonegawa1,Tonegawa2}. 
      The phase diagrams obtained by the NLSM are qualitatively 
agree with the numerical results. 

      In this paper we examine another mixed spin chain with 
period 4 which includes three species of spins. 
      The sequential order of magnitudes of spins in a unit cell 
is 1/2, 1, 3/2 and 1.  
      There are four kinds of nearest-neighbour exchange 
couplings among them. 
      If we consider a simplified case, the exchange 
parameters on the both sides of a 1/2 spin are the same value 
$J$, and those on the both sides of a 3/2 spin are of the same 
value $J'$. 
      We in fact consider the distortion of $J'$ 
on the both sides of a 3/2 spin. 
      We map the spin Hamiltonian describing the spin chain 
to an NLSM following Ref.~\cite{Takano1}. 
      From the value of the topological angle in the NLSM, 
we determine the phase diagram of this model in the 
space of the exchange parameters.

\section{Mapping to the nonlinear $\sigma$ model}

\begin{figure}[btp]
\begin{center}\leavevmode
\includegraphics[width=0.7\linewidth]{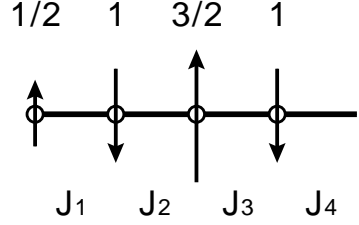}
\caption{ A unit cell of the 1/2-1-3/2-1 
spin chain. }
\label{Fig_cell}
\end{center}
\end{figure}

      The Hamiltonian for the present spin chain is written as 
\begin{eqnarray}
\label{spin_Hamiltonian}
      H &=& \sum_{j} 
 ( J_{1} \, {\bf S}_{4j+1} \cdot {\bf S}_{4j+2} 
 + J_{2} \, {\bf S}_{4j+2} \cdot {\bf S}_{4j+3} 
\nonumber \\
 &+& J_{3} {\bf S}_{4j+3} \cdot {\bf S}_{4j+4} 
 + J_{4} \, {\bf S}_{4j+4} \cdot {\bf S}_{4j+5} ) , 
\end{eqnarray}
where the magnitudes of ${\bf S}_{4j+1}$, ${\bf S}_{4j+2}$, 
${\bf S}_{4j+3}$ and ${\bf S}_{4j+4}$ are 
\begin{eqnarray}
\label{spin_magnitude}
(s_1, s_2, s_3, s_4) = 
\left( \frac{1}{2}, 1, \frac{3}{2}, 1 \right) . 
\end{eqnarray}
      A unit cell is illustrated in Fig. \ref{Fig_cell}. 
      To reduce the number of parameters, we restrict and 
reparametrize the antiferromagnetic exchange parameters as 
\begin{eqnarray}
\label{parametrization}
J_{1} = J_{4} = J , 
\quad 
J_{2} = \frac{J'}{1-\gamma} , 
\quad 
J_{3} = \frac{J'}{1+\gamma} , 
\end{eqnarray}
where $\gamma$ ($-1 < \gamma < 1$) is the distortion parameter 
describing the asymmetry between the couplings of the both sides 
of a spin with magnitude 3/2. 

      Following Ref. \cite{Takano1}, we can map a general periodic 
mixed spin chain satisfying a restriction to the NLSM. 
      The mapped NLSM action is given by 
\begin{eqnarray}
\label{action-NLSM}
      S_{\rm eff} = &{}& \int d\tau \int dx 
\biggl\{ 
- i \frac{J^{(0)}}{J^{(1)}} 
{\bf m} \cdot 
\left( \frac{\partial {\bf m}}{\partial \tau} \times 
\frac{\partial {\bf m}}{\partial x} \right) 
\nonumber \\ 
      &{}& + \frac{1}{2aJ^{(1)}} \left( 
\frac{J^{(1)}}{J^{(2)}} - \frac{J^{(0)}}{J^{(1)}} \right) 
\left( \frac{\partial {\bf m}}{\partial \tau} \right)^2 
\nonumber \\ 
      &{}& + \frac{a}{2} J^{(0)} 
\left( \frac{\partial {\bf m}}{\partial x} \right)^2 
\biggl\} ,
\end{eqnarray}
where ${\bf m}$ is an O(3) unit-vector field and $a$ is the 
lattice constant. 
      For a general spin chain with period $2b$, the constants 
in the action are given as follows: 
\begin{eqnarray}
\label{J_n_def}
\frac{1}{J^{(n)}} &=& \frac{1}{2b} \sum_{q=1}^{2b} 
\frac{({\tilde s}_q)^n}{J_{q} s_q s_{q+1}} 
\quad (n = 0, 1, 2) 
\end{eqnarray}
with accumulated spins 
\begin{eqnarray}
      {\tilde s}_q &=& \sum_{k=1}^{q} (-1)^{k+1} s_k . 
\quad (q = 1, 2, \cdots , 2b) 
\label{s_tilde_def}
\end{eqnarray}
      The action (\ref{action-NLSM}) is of the standard 
form of the NLSM: 
\begin{eqnarray}
\label{standard_NLSM}
       S_{\rm st} &=& \int d\tau \int dx 
\biggl\{ 
- i \frac{\theta }{4\pi} 
{\bf m} \cdot 
\left( \frac{\partial {\bf m}}{\partial \tau} \times 
\frac{\partial {\bf m}}{\partial x} \right) 
\nonumber \\ 
      &{}& + \frac{1}{2gv} 
\left( \frac{\partial {\bf m}}{\partial \tau} \right)^2 
+ \frac{v}{2g} 
\left( \frac{\partial {\bf m}}{\partial x} \right)^2 
\biggl\} , 
\end{eqnarray}
where $\theta$ is the topological angle, $g$ is the coupling 
constant and $v$ is the spin wave velocity. 

      In the present model, the accumulated spins 
(\ref{s_tilde_def}) are 
\begin{eqnarray}
\label{s_tilde_model}
({\tilde s}_1, {\tilde s}_2, {\tilde s}_3, {\tilde s}_4) = 
\left( \frac{1}{2}, -\frac{1}{2}, 1, 0 \right) 
\end{eqnarray}
and then the constants (\ref{J_n_def}) are calculated as
\begin{eqnarray}
\label{J_n_model}
&{}& \frac{1}{J^{(0)}} = \frac{1}{J} + \frac{1}{3J'} , 
\nonumber \\ 
&{}& \frac{1}{J^{(1)}} = \frac{1}{4} 
\left( \frac{1}{J} + \frac{1}{3J'} + \frac{\gamma}{J'} \right) , 
\nonumber \\ 
&{}& \frac{1}{J^{(2)}} = \frac{1}{8} 
\left( \frac{1}{J} + \frac{5}{3J'} + \frac{\gamma}{J'} \right) . 
\end{eqnarray}

\begin{figure}[btp]
\begin{center}\leavevmode
\includegraphics[width=1.0\linewidth]{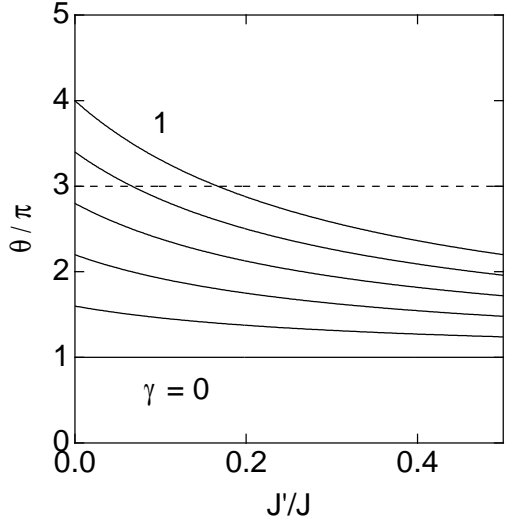}
\caption{ The topological angle $\theta$ as a function of 
$J'/J$ for $\gamma$ = 0.0, 0.2, 0.4, 0.6, 0.8 and 1.0. 
The system is gapless on the lines of $\theta/\pi$ = 1 and 
3 (dashed line).}
\label{Fig_angle}
\end{center}
\end{figure}

\begin{figure}[btp]
\begin{center}\leavevmode
\includegraphics[width=1.0\linewidth]{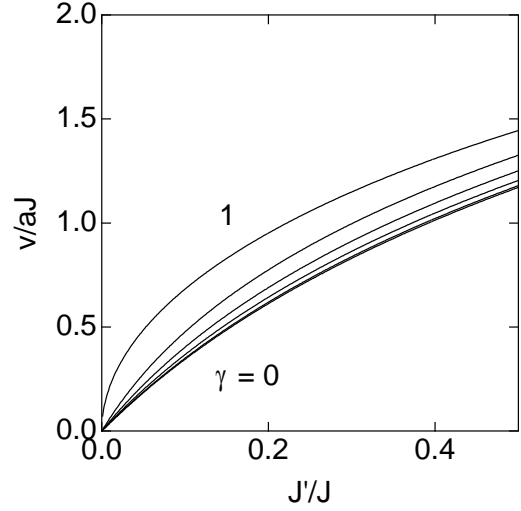}
\caption{ The spin wave velocity $v$ as a function of 
$J'/J$ for $\gamma$ = 0.0, 0.2, 0.4, 0.6, 0.8 and 1.0.}
\label{Fig_velocity}
\end{center}
\end{figure}

\begin{figure}[btp]
\begin{center}\leavevmode
\includegraphics[width=1.0\linewidth]{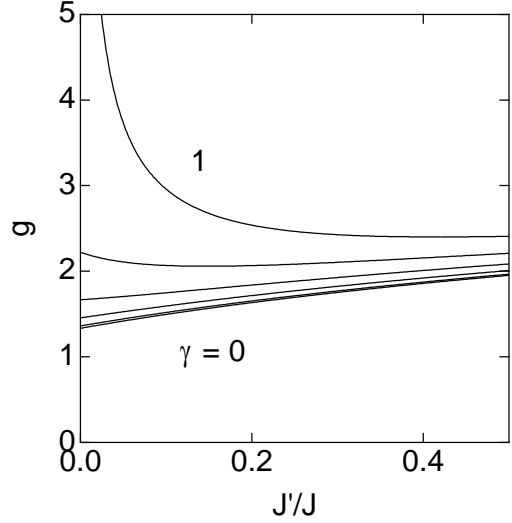}
\caption{ The coupling constant $g$ of the NLSM as a function 
of $J'/J$ for $\gamma$ = 0.0, 0.2, 0.4, 0.6, 0.8 and 1.0.}
\label{Fig_coupling}
\end{center}
\end{figure}

      Comparing Eq.~(\ref{action-NLSM}) with 
Eq.~(\ref{standard_NLSM}), we have the topological angle 
\begin{eqnarray}
\label{angle}
\theta = 4\pi \frac{J^{(0)}}{J^{(1)}} 
= \pi + \frac{\pi \gamma} 
{ \displaystyle\frac{J'}{J} 
     + \displaystyle\frac{1}{3} } , 
\end{eqnarray}
the spin wave velocity 
\begin{eqnarray}
\label{velocity}
v = 
\frac{4aJ'} 
{ \sqrt{
\left( \displaystyle\frac{J'}{J} 
     + \displaystyle\frac{1}{3} \right) 
\left( \displaystyle\frac{J'}{J} + 3 \right)
- \gamma^2 }
} 
\end{eqnarray}
and the coupling constant 
\begin{eqnarray}
\label{coupling}
g = 
\frac{4 \left( \displaystyle\frac{J'}{J} 
             + \displaystyle\frac{1}{3} \right)} 
{ \sqrt{
\left( \displaystyle\frac{J'}{J} 
     + \displaystyle\frac{1}{3} \right) 
\left( \displaystyle\frac{J'}{J} + 3 \right)
- \gamma^2 }
} . 
\end{eqnarray}
      These quantities are shown in Figs.~\ref{Fig_angle}, 
\ref{Fig_velocity} and \ref{Fig_coupling}, as functions of 
$J'/J$ for several values of $\gamma$. 

\section{Phase diagram}

      It is well known that an NLSM has gapless excitations 
if the topological angle $\theta$ is an odd-integer 
multiple of $\pi$. 
      In the general periodic spin model, this condition is 
turned into the {\it gapless equation} 
\begin{equation}
\frac{2J^{(0)}}{J^{(1)}} = \frac{2l-1}{2} 
\label{gapless_condition}
\end{equation}
with arbitrary integer $l$. 
      In the present case of Eq.~(\ref{angle}), 
it becomes of a simple form 
\begin{eqnarray}
\label{gapless_line}
\gamma = 
2 l \left( \frac{J'}{J} + \frac{1}{3} \right) . 
\quad (l = -1, 0, 1) 
\end{eqnarray}
      The equation for each $l$ determines a phase boundary 
between gapful disordered phases. 
      The restricted values of integer $l$ is due to the 
condition $-1 < \gamma < 1$. 
      In the symmetric case of $\gamma = 0$, the system is 
gapless irrespective of the ratio $J'/J$, because 
it satisfies Eq.~(\ref{gapless_line}) with $l = 0$. 

\begin{figure}[btp]
\begin{center}\leavevmode
\includegraphics[width=1.0\linewidth]{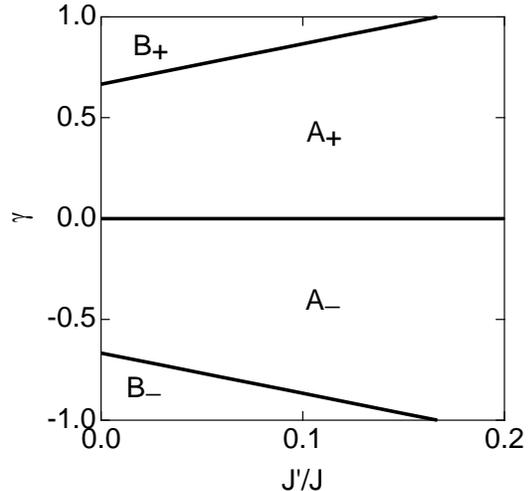}
\caption{ The phase diagram in the ($J'/J$, $\gamma$) 
parameter space ($-1 < \gamma < 1$).
  The phase boundaries are gapless lines determined by 
Eq.~(\ref{gapless_line}).  }
\label{Fig_Phase}
\end{center}
\end{figure}

      The phase diagram of the ground state is obtained 
by identifying the gapless lines (\ref{gapless_line}) as 
the phase boundaries. 
      We show the phase diagram in Fig.~\ref{Fig_Phase}. 
      There exist four phases A${}_+$, A${}_-$, B${}_+$ and 
B${}_-$ in the ($J'/J$, $\gamma$) parameter space. 
      The phase structure is symmetric for 
$\gamma \rightarrow - \gamma$ according to the symmetry of 
the Hamiltonian (\ref{spin_Hamiltonian}) with 
Eqs.~(\ref{spin_magnitude}) and (\ref{parametrization}). 

      In phase B${}_+$, $\gamma$ is close to 1 and $J_2$ 
is selectively large. 
      The ground state of B${}_+$ may be explained by a 
VBS picture \cite{Affleck2}. 
      That is, if we decompose each spin into spins with 
magnitude 1/2, the ground state is approximately a direct 
product of dimers. 
      Two dimers in a unit cell are formed between 1/2-spins 
at the both sides of a coupling with exchange parameter $J_2$. 
      Phase A${}_+$ is of an intermediate character and 
spreads over almost all area in the parameter space. 
      The ground state of A${}_+$ does not seem to be 
explained by the conventional VBS picture. 

\section{Summary}

      We have examined the 1/2-1-3/2-1 spin chain with a 
distortion of the exchange parameters by using the nonlinear 
$\sigma$ model method. 
      We obtained the phase diagram of the ground state in 
the parameter space. 
      There exist four gapful phases. 
      In the case of no distortion, the system is always 
gapless and is on a phase boundary. 
      When the distortion is strong, there are two phases 
to be explained by a VBS picture. 
      It is a future work to explain the other phases 
by the singlet-cluster-solid (SCS) picture developed in 
Ref. \cite{Takano3}.


\end{document}